\documentstyle[11pt,epsfig,amssymb]{article}
\newlength{\dinwidth}
\newlength{\dinmargin}
\newcommand {\nn} {\nonumber}
\newcommand {\half} {\frac{1}{2}}
\newcommand {\p} {\prime}
\newcommand {\G} {{\cal G}}
\newcommand {\M} {{\cal M}}

\newcommand{\beq}{\begin{equation}}
\newcommand{\eeq}{\end{equation}}
\newcommand{\beqa}{\begin{eqnarray}}
\newcommand{\eeqa}{\end{eqnarray}}

\newcommand{\fr}{\frac}
\newcommand{\mn}{\mu \nu}
\newcommand{\nab}{\bar{\nabla}}
\newcommand{\R}{\bar{R}}
\newcommand{\ta}{\tilde{\alpha}}
\newcommand{\tb}{\tilde{\beta}}
\newcommand{\g}{\bar{g}}
\newcommand{\wt}{\widetilde}
\newcommand{\wwh}{\widehat{\hat{h}}}
\newcommand{\wwp}{\widehat{\hat{\pi}}}
\renewcommand{\theequation}{\arabic{equation}}

\begin{document}
\thispagestyle{empty}
\addtocounter{page}{-1}
\begin{flushright}
SNUTP-26/01\\
{\tt hep-th/0108194}
\end{flushright}
\vspace*{1.3cm}
\centerline{\Large \bf Effective Lagrangian from Higher Curvature Terms:}
\vskip0.4cm
\centerline{\Large \bf Absence of vDVZ Discontinuity in AdS Space}
\vspace*{0.8cm}
\centerline{\bf  Ishwaree P. Neupane\,\footnote{\tt \quad E-mail address:
ishwaree@phya.snu.ac.kr}}
\vspace*{0.4cm}
\centerline{\it School of Physics, Seoul National University, 151-742,
Seoul, Korea}
\vspace*{0.3cm}
\vspace*{1.0cm}
\centerline{\bf Abstract}
\vspace*{0.4cm}
We argue that the van Dam-Veltman-Zakharov discontinuity
arising in the $M^2 \to 0$ limit of the massive graviton through
an explicit Pauli-Fierz mass term could be absent in anti de Sitter
space. This is possible if the graviton can acquire mass spontaneously
from the higher curvature terms or/and the massless limit $M^2\to 0$ is
attained faster than the cosmological constant $\Lambda \to 0$.
 We discuss the effects of higher-curvature couplings and of an explicit
cosmological term ($\Lambda$) on stability of such continuity and of
massive excitations.

\begin{flushleft}
\vspace*{0.2cm}

\end{flushleft}

\vspace*{0.6cm}

\baselineskip=18pt

\newpage

\setcounter{equation}{0}

\section{Introduction}
Has the graviton exactly zero-mass or perhaps
a small but non-zero mass? This is somehow related to whether or not the
$M^2\to 0$ limit in a massive gravity corresponds to a strictly massless
graviton, with $M^2$ being the mass of a spin-$2$ excitation. In the
massive gravity theory defined by the Pauli-Fierz~\cite{FP}
term, one finds a discrete difference between the propagators for
the massless graviton and that for the graviton with mass $M^2\to 0$.
Recently this has got a new impetus in an
anti de Sitter ($AdS$) space~\cite{Kogan,Porrati}. The issue is subtle and
appealing, and one has to resolve this by looking at the massive gravity
theories defined by other than the Pauli-Fierz action. A key word of
this approach is that van Dam-Veltman-Zakharov (vDVZ)
discontinuity\cite{DVZ} known to exist in Minkowski space
may not survive in $AdS$ space, in particular, if a smooth limit
$M^2/\Lambda \to 0$ exists. Kogan et.al. in~\cite{Kogan} and Porrati
in~\cite{Porrati} used the spin-2 Pauli-Fierz mass term to
demonstrate the absence of vDVZ discontinuity in $AdS$ space at
tree level, and further realization of this non-perturbative
continuity for the spin $3/2$ field of ${\cal N}=1\, AdS$
supergravity has been given in~\cite{GvN,Deser}. Absence of vDVZ
discontinuity in de Sitter (dS) space was already shown by
Higuchi~\cite{Higuchi}, but as the result was restricted there to de
Sitter space and hence may have less importance, for there is no
unitary spin-$2$ representation in the mass range $0<M^2<2\Lambda/3$.
A smooth limit $M^2/\Lambda\to 0$ has been observed in~\cite{KR} for the
four-dimensional theory of massive AdS gravity by embedding it into five
dimensional AdS brane-world models.

In order to achieve a smooth limit $M^2/\Lambda\to 0$, perhaps one should
find a mass term $M^2$ of the order $\Lambda^2$, which could arise
from the terms that are quadratic in the curvature. Naively, one finds
reasonable to introduce the higher curvature terms in a ghost free
combination into the starting action. A concrete example is the
Gauss-Bonnet (GB) term~\cite{Zumino,DGB}, which is a ghost-free
combination~(see Ref.~\cite{IPN3} for the absence of ghost state in
$AdS_5$ based a single brane model). But it is obvious that the GB term
does not generate a mass term for graviton in four-dimensional flat or
curved spaces. Eventually, one could attempt the problem by introducing
the higher-curvature terms in generic form. In this paper we explore this
possibility and find a smooth limit for $M^2/\Lambda\to 0$, where the
$M^2$ could be quadratic one of $\Lambda$.
It is possible that a very light graviton in the presence of non-zero
cosmological constant does not show a significant deviation from the
observational effects that of massless gravitons at large distance scales.
For example, the prediction of Einstein's gravity for the bending of light
by the Sun agrees to the observation with less than $1\%$ difference,
a limit $M^2/\Lambda\leq 10^{-2}$, therefore, must be satisfied by the
graviton mass $M^2$, and hence a small graviton mass may not be in
contradiction with the observation~\cite{Kogan2,Dvali}.

An undeniable fact is that the starting gravitational theory is
Einstein gravity, thus any natural modification on it should arise from
the higher order corrections in the curvature. Among the different
motivations in higher derivative theory, the special one here is attributed
to an expectation that mass term arises from terms that are quadratic in
the curvature, so that a smooth $M^2/\Lambda\to 0$ limit can exist. It is
really astonishing that four-derivatives gravitation in a curved background
that we study here is a crucial tool to show the absence of
van Dam-Veltman-Zakharov discontinuity in an AdS space. Without any
ambiguity we understand that the higher derivative theory with terms upto
quadratic in the curvatures effectively contains, which also we explicitly
demonstrate in this paper, in addition to
the usual massless graviton excitation, a massive spin-two ghost and
a massive scalar. Nonetheless, by judiciously choosing the
higher derivative coupling $\tb~(\equiv \beta+4\gamma)$ zero, one
can always recover a theory which is ghost free. The most radical
result is that one recovers the correct expressions for the
massless and the massive graviton propagators by taking simultaneously
the limits $\tb\to 0,~\Lambda\to 0$ but $\Lambda>> \tb$, and
$\Lambda\to 0,~\tb\to 0$ but $\Lambda<< \tb$. We also present several
novel results in a flat spacetime background.

The outline of the paper is as follows. In Section 2 we give some
basic features of the fourth-order gravity in a curved
background showing that one generates a massless spin-2 graviton and extra
massive degrees of freedom with the higher derivatives. In Section 3
we evaluate the one-particle exchange amplitude between two conserved
sources and demonstrate the absence of vDVZ discontinuity in the
$\Lambda\to 0$ limit. In Section 4 we briefly discuss the
issue of the general covariance with a (non-)trivial $\Lambda$. We also
briefly describe the spin and field content of the linearized
four-derivative equations, and the counting of degrees of freedom by
introducing gauge fixing term. Section 5 consists of conclusions. Finally,
Appendix A presents the basic variational formulae for different curvature
terms.

\section{Effective action and Linear equations}

We start with the Einstein-Hilbert action supplemented with the quadratic
curvature terms:
\beq S_G = \int d^4x \sqrt{- g} \Big \{ \fr{1}{\kappa^2}\, R
-2\Lambda + \alpha R^2 +\beta  R^2_{\mn} + \gamma
R^2_{\mu\nu\rho\sigma}\Big \}+ S_{gf}\,. \label{action}
\eeq
Here $S_{gf}$ is the gauge fixing action. Here $\kappa^2
\alpha,\,\kappa^2\beta,\,\kappa^2 \gamma$ have dimensions of
inverse mass squared, and $\kappa^2 \Lambda$ has dimension $M^{2}$.
For simplicity, we set $\kappa^2= 16 \pi
G_4=1$, unless otherwise stated. We may keep $\Lambda$ arbitrary,
but most of the results bear meaningful interpretations for
$\Lambda<0$ or $\Lambda=0$. It is known that the theory
defined by~(\ref{action}) explains, in addition to the usual massless
graviton, two new fields: a massive spin-$2$ field with five degrees of
freedom (DOF), and a massive scalar with one DOF~\cite{KStelle,Hindawi}.

The equations of motion derived from~(\ref{action}) read as~\cite{IPN1},
in units $\kappa^2=1$,
\beqa
&&\left(R_{\mn}-\fr{1}{2}g_{\mn}R\right)+
\Lambda g_{\mu\nu}+2\alpha R\left(R_{\mu\nu}-\frac{1}{4}
g_{\mu\nu} R\right)+2\beta\left(R_{\mu\rho\nu\sigma}
R^{\rho\sigma}
-\frac{1}{4}g_{\mu\nu} R_{\rho\sigma}R^{\rho\sigma}\right)+\nn \\
&&~~~~~~~+2\gamma \left(R_{\mu\rho\sigma\delta} R_\nu\,^{\rho\sigma\delta}
-\frac{1}{4}\,g_{\mu\nu} R_{\rho\sigma\delta\lambda}
R^{\rho\sigma\delta\lambda} -2 R_\mu\,^\lambda R_{\nu\lambda}
+2R_{\mu\lambda\nu\rho} R^{\lambda\rho}\right)-\nn \\
&& ~~~~~~~-(2\alpha+\beta+2\gamma)\left(\nabla_\mu
\nabla_\nu-g_{\mu\nu} \nabla^2\right)R +(\beta+4\gamma)\,
\nabla^2\left(R_{\mu\nu}-\frac{1}{2}\, g_{\mu\nu} R\right)=0
\label{eq.of.mo}
\eeqa
In the linear approximations, the curvature terms like $R R_{\mu\nu},\,
R_{\mu\rho\nu\sigma} R^{\rho\sigma}$ affect the gravitational excitation
spectrum near flat space. That is, in a flat spacetime background
($\Lambda=0$), second, third and fourth bracket terms in~(\ref{eq.of.mo})
do not contribute to the equations of motion linear in fluctuation, but they
do contribute in the curved backgrounds. The $AdS$ (or $dS$)
space solution is expressed in terms of the background metric
($\bar g_{\mn})$  as
\beq
\R_{\mu\nu\rho\sigma}=\fr{\Lambda}{3}(\bar{g}_{\mu\rho}\bar{g}_{\nu\sigma}
- \bar{g}_{\mu\sigma}\bar{g}_{\nu\rho}),~~~ \R_{\mu\nu}= \Lambda
\bar{g}_{\mu\nu},~~~ \R=4 \Lambda\,. \label{Sol}
\eeq
Here $\Lambda=-3/l^2$ is defined, where $l$ is the $AdS_4$ length scale.
It is trivial to check that the
background solution~(\ref{Sol}) also solves the zeroth order equations of
motion (e.o.m.) of higher derivative gravity defined by~(\ref{action}).
One must replace $g_{\mu\nu}$ in~(\ref{eq.of.mo}) by $\bar{g}_{\mu\nu}$.
Of course, the covariant derivatives
taken about the constant (negative) curvature backgrounds are vanishing.
A field
satisfying $\bar{R}_{\mu\nu}=\Lambda \bar{g}_{\mu\nu}$, $\bar{R}=4\Lambda$
always solves~(\ref{eq.of.mo}). The sum of the first two terms
(i.e. $\bar{G}_{\mu\nu}+\Lambda\bar{g}_{\mu\nu}$) in~(\ref{eq.of.mo}) gives
zero. Also note that second, third, and fourth round bracket term
in~(\ref{eq.of.mo}) taken about the background~(\ref{eq.of.mo}), is each
separately vanishing, hence one finds the standard e.o.m. of massless
gravity for any value of $\Lambda$. This is completely a consistent
argument, because unless we perturb the background, we do not except any
massive (tensor or scalar) modes to be present.

In order to study the propagation of the metric and the complete mass
spectrum in four-dimensions, we
introduce the perturbation around the background space $g_{\mu\nu} =
\bar{g}_{\mu\nu} +h_{\mu\nu}$. Then to the order linear
in $h_{\mu\nu}$, neither the derivatives terms, nor the bracket terms
in~(\ref{eq.of.mo}) are vanishing in the curved backgrounds. Note that
the source term for the metric
fluctuation $h_{\mu\nu}$ can then be defined by
\beq S_M= - 2\int
d^4x\,\sqrt{-g}\,h_{\mu\nu} T^{\mu\nu}\,.
\eeq
We can further make a simplifying assumption that $T_{\mu\nu}$ is
co-variantly conserved with respect to the background metric, $\nab _\mu
T^{\mu\nu}=0$. For the study of massive gravity, a common practice in the
literature~[1-7,17] is that one adds the following Pauli-Fierz action to
the linearized action for $h_{\mu\nu}$ derived from~(\ref{action}) (but
setting $\alpha=\beta=\gamma=0$)
\beq
S_{PF}=M_{PF}^2\int d^4x\,\sqrt{-g}
\Big[h_{\mu\nu} h^{\mu\nu}-
\big(h_\lambda\,^\lambda\big)^2\Big]\,,\label{Pau-Fie}
\eeq
where $M_{PF}^2$ is a pure spin-$2$ mass term.
A clear motivation for introducing this term is that it is
a ghost-free term for a free, massive spin-2 propagation at the
linearized level. However, as the mass is acquired in an explicit way,
one cannot, in general, expect a smooth limit at the quantum
level~\cite{Duff}. Remarkably, we find that~(\ref{Pau-Fie}) is not
essential to prove the absence of discontinuity in graviton mass in an
$AdS$ background. Beside, we have no good reason yet why a spin-$2$
excitation should have an arbitrary real mass in the pure four-dimensional
gravity. Indeed, in a pure $4d$ theory, addition of the leading order
quadratic curvature terms to the Einstein-Hilbert action (the total action
then explains a massive or multi mass model of gravity) is more logical
rather than an addition of the explicit Pauli-Fierz (PF) mass term.
Besides, a purely four-dimensional theory of massive gravity with the
PF term is not well defined in four-dimensions~\cite{Dvali}, because an
arbitrary spin-$2$ field does not guarantee that this field will couple
to a conserved source. The origin of this term is expected to arise
from some massive graviton sector of the usual Kaluza-Klein reduction
(see for example Ref.~\cite{YMC} for a close realization). We simply
the drop the Pauli-Fierz action in the present analysis.

In order to obtain linear equations of motion for $h_{\mu\nu}$ we use the
following relations
\begin{eqnarray}
\delta R_{\mu\nu\alpha}\,^{\beta}&=& -\nab_{[\mu}\nab_{|\alpha|}
h_{\nu]}\,^\beta +\nab_{[\mu}\nab^\beta h_{\nu]\,\alpha}
+\half\left(\R_{\mu\nu\rho}\,^\beta h_\alpha\,^\rho- R_{\mu\nu\alpha}\,^\rho
h_\rho\,^\beta\right)\label{curvature} \\
2\delta R_{\mu\nu}&=&\Delta_L^{(2)} h_{\mu\nu}
-\nab_\mu\nab_\nu h
+2\nab_{(\mu}\nab^\rho h_{\nu)\rho}\,,\label{Riccitensor}\\
\delta R &=& -\bar{R}_{\sigma\rho} h^{\sigma\rho}- \bar{\nabla}^2 h  +
\bar{\nabla}_\sigma\bar{\nabla}_\rho h^{\sigma\rho}\,,\label{Ricciscalar}
\end{eqnarray}
where the Lichnerowicz operator $\Delta_L^{(2)}$ acting over the
field $h_{\mu\nu}$ is given by
\beq
\Delta_L^{(2)}
h_{\mu\nu}= -{\nab}^2h_{\mn} -2 \R_{\rho\mu\alpha\nu}
h^{\rho\alpha} + 2\R_{\rho(\mu}h^\rho~_{\nu)}=-\nab^2 h_{\mu\nu}
+\frac{8\Lambda}{3}\left(h_{\mu\nu}-\frac{1}{4}\,\g_{\mu\nu} h\right)\,,
\eeq
and the operators $\Delta_L$'s further obey the following
properties~\cite{MJDuff}
\beqa
&&\nab^\mu\Delta_L^{(2)}h_{\mu\nu}=\Delta_L^{(1)}\nab^\mu h_{\mu\nu}\,, ~
\nab^\mu\Delta_L^{(1)}f_\mu=\Delta_L^{(0)}\nab^\mu f_\mu\,,\nn\\
&&\Delta_L^{(2)}\g_{\mu\nu}h=-\g_{\mu\nu}\nab^2 h\,,
~ \Delta_L^{(1)}f_\mu=\left(-\nab^2+\Lambda\right)f_\mu\,.
\eeqa
Here $\Delta^{(1)}_L$ and $\Delta^{(0)}_L$ are the Lichnerowicz
(differential) operators acting respectively on the spin-$1$ and
spin-$0$ fields. Note that $\Delta_L^{(2)}\,,\Delta_L^{(1)}$ and
$\Delta_L^{(0)}$ commute with
contraction and covariant differentiation about the background. We can
easily show that the local curvature squared terms in the GB combination
does not generate mass for the linear perturbations. This is because
in ordinary two derivative gravity supplemented by
the GB term, each of the four gauge-group local parameters of
Diff-invariance accounts for killing of two degrees of freedom (DOF),
leaving just two DOF of the massless graviton out of ten DOF of
$h_{\mu\nu}$. A general feature of gravitational action
with arbitrary $\alpha, \beta, \gamma$ is, however, that the
graviton excitations near a flat space could behave as ghosts. This
means that the set of all states would not form a Hilbert space with a
positive definite metric. However, it is worth noting that the theory
defined by~(\ref{action}) is ghost free when a relation
$\beta=-4\gamma$ holds.

To the first order in $h_{\mu\nu}$ and up to the terms that vanish on the
Einstein backgrounds~(\ref{Sol}) (see Appendix A for the details),
the linear equations of motion in $D=4$ take the following form
\beqa
&&(1+a_1)\Delta^{(2)}_Lh_{\mu\nu} +
2(1+a_2)\nab_{(\mu}\nab^\rho h_{\nu)\rho}
 -(1+a_3)\nab_{(\mu}\nab_{\nu)}h +2\Lambda h_{\mu\nu}\nn \\
&&-4(1+a_3)\Lambda\big(h_{\mu\nu}-\frac{1}{4}\g_{\mu\nu} h\big)
+(1+a_4) \g_{\mu\nu}\nab^2h-(1+a_5) \g_{\mu\nu}\nab_\lambda\nab_\rho
h^{\lambda\rho}\nn \\
&&~~~~~~~~+2(2\ta+\tb)\big(\nab_\mu\nab_\nu-\g_{\mu\nu}\nab^2\big)\,
\big(\nab^2 h-\nab_\lambda\nab_\rho h^{\lambda\rho}\big)
=4 T_{\mu\nu}\,,
\label{neweqn}
\eeqa
where $\nab_\mu $ is taken to be covariant with
$\nab _\mu \bar{g}_{\lambda\rho}=0$,~ $\ta=\alpha-\gamma\,, ~
\tb=\beta+4\gamma$, and the coefficients $a_i$ are defined by
\beqa
&&a_1=8\Lambda \ta +\tb \Big(\nab^2-\frac{8\Lambda}{3}\Big)\,, ~~
a_2=8\Lambda\ta +\tb \Big(\nab^2+\frac{4\Lambda}{3}\Big)\,,\nn \\
&&a_3=4\Lambda\ta +\tb \Big(\nab^2-\frac{2\Lambda}{3}\Big)\,, ~~
a_4=\tb \Big(\nab^2-\frac{4\Lambda}{3}\Big)\,,\nn \\
&& a_5=4\Lambda\ta +\tb \Big(\nab^2+\frac{2\Lambda}{3}\Big)\,.
\label{coeffs}
\eeqa
The trace of Eq.~(\ref{neweqn}) takes the following form
\beq
\left[2-4\tb\left(\nab^2+\frac{\Lambda}{2}\right)-12\ta
\left(\nab^2+\Lambda\right)\right]\nab^2 h -\left[2-4\tb \nab^2-12\ta
\nab^2\right] \nab^\mu\nab^\nu h_{\mu\nu}+2\Lambda h=4T\,.
\label{trace}
\eeq
An obvious consequence of fourth-order derivatives of the metric in
field equations is that there are six more degrees of freedom in the
theory other than the two massless graviton states~\cite{KStelle}. One may
also define a condition to constrain the number of propagating degrees
of freedom. A physical choice is $\nab^\mu\nab^\nu h_{\mu\nu}=\nab^2 h$,
which indeed satisfies the field decompositions given in the
Ref.~\cite{KStelle}. It is worth noting that in the massive
gravity theory given by the Pauli-Fierz(PF) action, this constraint
naturally appears from the field equations~\cite{Porrati}, which simply
follows by taking a double divergence of the linearized equation containing
$M_{PF}^2(h_{\mu\nu}-\bar{g}_{\mu\nu} h)$. Since the
covariant derivative of the Einstein-Hilbert term is zero, one finds
$M_{PF}^2(\nab^\nu h_{\mu\nu}-\nab_\nu h)=0$. This further implies that
$\nab^\mu\nab^\nu h_{\mu\nu}=\nab^2 h$ for on-shell $h_{\mu\nu}$,
and as we see below this is more useful to compute the one-particle
amplitude.

Then from the Eq.~(\ref{trace}), we obtain
\beq
\big(3{\cal M}^2-2\Lambda\big) h=-4T\,,\label{masstermeqn}
\eeq
where we have defined
${\cal M}^2 = 2\Lambda\left(\tb+6\ta\right)\nab^2/3$.
Meanwhile, if we choose $\tb+3\ta=0$ in~(\ref{trace}), we find
\beq
-2\nab^2 h+2\nab^\mu\nab^\nu h_{\mu\nu}-2\Lambda h +3\M^2 h=-4T \,.\label{po-eqn}
\eeq
By replacing ${\cal M}^2$ with $M_{PF}^2$, one observes that the
Eqs.~(\ref{masstermeqn}) and~(\ref{po-eqn}) become identical with
the Eqs.(13) and (11) of the Ref.~\cite{Porrati}. Nevertheless, there
is a clear difference in their origins, and ${\cal M}^2$ defined above is not
what a PF mass term is. $M_{PF}^2$ is a pure mass
term put by hand, while $\M^2$ is just like a mass parameter, whose value is
determined only when one defines the pole $\nab^2$. When we express the
curvature terms in the GB combination, we have $\ta=\tb=0$, and
hence ${\cal M}^2=0$. This is expected because the GB combination is a
topological in four-dimensions and does not change the Einstein's gravity.

To get something in concrete form from the Eqs.~(\ref{neweqn})
and~(\ref{trace}), we need to deploy the following relations
\beqa
&&\nab_\mu\left(\nab_\nu\nab_\lambda+\frac{\Lambda}{3}\,g_{\nu\lambda}\right) f
=\nab_\nu\left(\nab_\mu\nab_\lambda+\frac{\Lambda}{3}\,g_{\mu\lambda}
\right)f \\
&&\nab^2\left(\nab_\mu\nab_\nu+\frac{2\Lambda}{3}\,g_{\mu\nu}\right)f =
\nab_\mu\nab_\nu\left(\nab^2+\frac{8\Lambda}{3}\right)f \\
&&\nab^\mu\left(\nab_\mu f_\nu+\nab_\nu
f_\mu\right)=-\Delta_L^{(1)} f_\nu +2\Lambda f_\nu +\nab_\nu
\nab^\mu f_\mu\,. \eeqa Apply $\nab^\mu$ to Eq.~(\ref{neweqn}).
Using the above relations we find
\beq
\Lambda\tb\left(\nab^2-11\Lambda\right) \nab^\mu h_{\mu\nu}
-\Lambda\tb \nab_\nu \nab^\mu\nab^\sigma h_{\mu\sigma}=0\,.
\label{singlediv.}
\eeq
For instance, if we impose $\nab^\mu\nab^\sigma
h_{\mu\sigma}=0$, then Eq.~(\ref{singlediv.}) implies that \beq
\Lambda\tb \left(\nab^2-11\Lambda\right)\nab^\mu h_{\mu\nu}=0\,.
\label{tracecondition}
\eeq
Here if $\Lambda\tb \left(\nab^2-11\Lambda\right)$ is defined by some mass
parameter $m^2$, the Eq.~(\ref{tracecondition}) reveals that only $10-4=6$
(Not $5!$) components of $h_{\mu\nu}$ propagate independently. Five of these
correspond to a massive spin-$2$ particle and the sixth
corresponds to a massive scalar. However, with
$\nab^\mu\nab^\nu h_{\mu\nu}=\nab^2 h$, we can further reduce the number
to the five DOF of a massive spin-$2$ field, which corresponds to a
ghost-like particle. But we can make theory free from ghost by choosing
$\tb=0$, which is the choice we would like to have in most of the discussion.

Finally, compute the double divergence of Eq.~(\ref{neweqn}) (or simply
apply $\nab^\nu$ to Eq.~(\ref{singlediv.})) and arrive to
\beq
-10 \Lambda^2 \tb \nab^\mu\nab^\nu h_{\mu\nu}=0\,.
\eeq
As we see below, for the absence of an unphysical pole, we must take
$|\Lambda\tb|=0$, so the possible choices are $\Lambda =0,~\tb\neq 0$;
$\Lambda\neq 0,~\tb=0$; $\Lambda=\tb=0$. The choice $\tb=0$ is more
preferred because it makes the theory free from the massive
spin-$2$ ghost, and also makes the residue at an unphysical
pole zero. Otherwise the equality $\nab^\mu\nab^\nu h_{\mu\nu}=0$ should
hold.

\section{One-particle amplitude}
Now we decompose $h_{\mu\nu}$ as follows
\beq \label{decompo}
h_{\mu\nu}=h_{\mu\nu}^{TT}+ \nab_{(\mu}
V_{\nu)}+\nab_\mu\nab_\nu\phi +g_{\mu\nu}\psi\,.
\eeq
This, with the conditions $\nab_\mu V^\mu=0$ and $\nab^\mu h_{\mu\nu}^{TT}
=0=h^{TT}$, further implies that
\beq
\nab^2h=\nab^4\phi+4\nab^2\psi\,, ~~
\nab^\mu\nab^\nu h_{\mu\nu}=\nab^4 \phi +\Lambda \nab^2\phi
+\nab^2 \psi\,.\label{hphipsi}
\eeq
The one-particle amplitude is most easily computed by choosing
$\nab^\rho \nab^\lambda h_{\lambda\rho}=\nab^2 h$, thus
from~(\ref{hphipsi}) we find $3\nab^2\psi=\Lambda\nab^2\phi$.
From~(\ref{decompo}), we then easily
read off $h=(3\nab^2+4\Lambda)\psi/\Lambda$. Use this value of $h$ in
Eq.~(\ref{masstermeqn}) and arrive to
\beq
\left(3\nab^2+4\Lambda\right) \psi= -4\Lambda
\left(2\Lambda\left(\tb+6\ta\right)-2\Lambda\right)^{-1} \,T
\eeq The transverse traceless
(TT) part of Eq.~(\ref{neweqn}) is
\beq
\left((1+a_1)
\Delta_L^{(2)}-2\Lambda (1+2a_3)\right)h_{\mu\nu}^{TT} =4
T_{\mu\nu}^{TT}\,,
\eeq
where the TT-component of $T_{\mu\nu}$ can
be expressed in the following form
\beq
T_{\mu\nu}^{TT}=T_{\mu\nu}-\frac{1}{3}\,g_{\mu\nu} T
+\frac{1}{3}\left(\nab_\mu\nab_\nu+g_{\mu\nu}\Lambda/3\right)\,
\left(\nab^2+4\Lambda/3\right)^{-1}\, T\,.
\eeq
Then the one-particle exchange amplitude between two co-variantly conserved
sources $T^{\prime}_{\mu\nu}$ and $T_{\mu\nu}$ can be simply
written as
\beqa
A&=&\half \int d^4x
\sqrt{-g}\,{T'}^{\mu\nu}(x)\,h_{\mu\nu}(x)
\equiv \half\, {T'}^{\mu\nu} h_{\mu\nu}^{TT}+\half T' \psi\nn \\
&=& 2\, {T'}^{\mu\nu}\frac{1}{(1+a_1)\Delta_L^{(2)}
-2\Lambda(1+2a_3)}\,T_{\mu\nu}-\frac{2}{3}\, T'\,
\frac{1}{-(1+a_1)\nab^2-2\Lambda (1+2a_3)}\,T\nn \\
&{}& + \frac{2\Lambda}{9}\,
T'\,\frac{1}{\left(-(1+a_1)\nab^2-2\Lambda
(1+2a_3)\right)\,\left(\nab^2+4\Lambda/3\right)}\,T\nn \\
&{}&-\frac{2\Lambda}{3}\,T'\,
\frac{1}{2\Lambda\left((\tb+6\ta)\nab^2-1\right)\,
\left(\nab^2+4\Lambda/3\right)}\,T\,.\label{amplitude}
\eeqa
Obviously, there are four propagators that may have poles in the
amplitude Eq.~(\ref{amplitude}), and we have to find the residue
at each pole. We now proceed to study each and every poles.

\section*{\it (A) Pole at $\nab^2=-4\Lambda/3$}

The residue at the unphysical pole $\nab^2=-4\Lambda/3$ is given by
\beqa
&&\left(\frac{2\Lambda}{9}\right)\,\frac{-3}{2\Lambda+8\Lambda^2(2\ta
-\tb)} -\left(\frac{2\Lambda}{3}\right)\,
\frac{3}{-6\Lambda-8\Lambda^2(6\ta+\tb)}\nn \\
&&~~~~~~~~~~~~~~~~~~~~~~~~~~~~~~ = - \frac{16\Lambda\tb}
{3\,\left[1+4\Lambda(2\ta-\tb)\right]\,\left[3+4\Lambda(6\ta+\tb)\right]}\,.
\label{residue} \eeqa
Hence the residue can be made zero only by taking $\Lambda=0$ or $\tb=0$,
these are the choices we would like to have henceforth.

\section*{\it (B) Pole at $\nab^2=-\frac{2\Lambda(1+2a_3)}{(1+a_1)}\equiv
Q^2$}

As stated above, for a meaningful result, the residue at the unphysical
pole should to zero, which is achieved by taking $\Lambda=0$ or $\tb=0$.
The propagator structure, however, depends upon the fact that out of
$\tb$ and $\Lambda$ which one vanishes first. Thus we find two
possibilities.

\subsection*{(I) $|\Lambda|>>\tb$}

By setting $\tb =0$ first, one kills the massive spin-$2$ degrees of
freedom, and hence the usual massless spin-$2$ graviton propagator can
be recovered. By substituting the values of
$a_1$ and $a_3$ from~(\ref{coeffs}), we obtain
\beqa
\nab ^2&=&-\frac{2\Lambda(1+a_3)}{(1+a_1)}=
\frac{-2\Lambda\,\left[1+8\Lambda\ta+
2\tb \left(\nab^2-2\Lambda/3\right)\right]}
{1+8\Lambda\ta+\tb\left(\nab^2-8\Lambda/3\right)}\nn \\
&{}& \stackrel{\tb= 0}
{\longrightarrow}-2\Lambda_{eff}+M^2_{eff}=-2\Lambda\,,
\eeqa
where
\beq
\Lambda_{eff}=\frac{\Lambda}{1+8\Lambda \ta}\,, ~~
M_{eff}^2=\frac{2{\cal M}^2}{1+8\Lambda \ta}\,, ~~
{\cal M}^2=-8\ta \Lambda^2 >0\,, ~~~
(\ta<0\,, ~ |\Lambda \ta|<<1)\,.
\eeq
The residue at $\nab^2=Q^2$ is
\beq
\left[-\frac{2}{3}+\frac{2\Lambda}{9}\cdot \left(\frac{-3}{2\Lambda}\right)
\right]\,\frac{1}{1+8\Lambda \ta}=-\frac{1}{1+8\Lambda \ta} ~ \,.
\eeq
In the limit $\Lambda \to 0$\,, i.e. $\M^2\to 0$, the residue is always $-1$,
so that
\beq
A=\lim_{\Lambda\to 0}\, \frac{G_N}{(1+8\Lambda \ta)}\,
\left[2{T^\prime}^{\mu\nu}\, \wt{\Box}^{-1} \,T_{\mu\nu}-
T'\,\wt{\Box}^{-1}\,T\right]\,,
\eeq
where $\wt{\Box}=-\nab^2-2\Lambda_{eff}+M_{eff}^2$. It might be relevant to
compare this to the physical pole at $\nab^2=M_{PF}^2-2\Lambda$ when one
introduces an explicit Pauli-Fierz mass term instead of higher derivative
terms. When $\ta\to 0$, $\Lambda_{eff}\to \Lambda$ and $M^2_{eff}\to 0$.
Further, if we set $\Lambda=0$, one finds the well known one-particle
exchange amplitude for a massless graviton in the pure Einstein's theory. This
may be enough to understand the absence of vDVZ discontinuity in an
$AdS$ space. As expected, for $\tb=0$, we find a massless graviton of the
Einstein-Hilbert theory, but the $4d$ Newton constant can be renormalized
due to the non-trivial $\ta$. Since all the extra degrees of freedom decouple
for $\tb= 0$, there is no any ghost state in the theory, and also
that the residue at an unphysical pole mentioned above is vanishing.

\subsection*{(II) $|\Lambda|<<\tb$}

For general $\ta$ and $\tb$, the two physical poles at $\nab^2=Q^2$ imply
the following quadratic equation
\beq
\tb\nab^4+\left(1+8\Lambda\ta+\frac{4\Lambda\tb}{3}\right)\nab^2
+2\Lambda\left(1+8\Lambda\ta-\frac{4\Lambda\tb}{3}\right)=0\,.
\eeq
Clearly, there are two poles corresponding to positive and negative
roots of $\nab^2$ (with $|\ta\Lambda|<<1$ and $|\tb\Lambda|<<1$)
\beqa
&&\nab_-^2=-2\Lambda+\frac{8\Lambda^2}{3}\left(\tb-6\ta\right)
\equiv -2\Lambda+{\cal M}_1^2\,,\label{nabbla-minus}\\
&&\nab_+^2= -\frac{4\Lambda}{3}-\frac{\left(1+8\Lambda\ta\right)}{\tb}
\equiv -\frac{4\Lambda}{3}-{\cal M}_2^2\,,\label{nabla-plus}
\eeqa
If $\Lambda$ goes to zero first, we have ${\cal M}_1^2=0$ and
${\cal M}_2^2=1/\tb$. If we further send $\tb\to 0$, the massive spin-$2$
field decouples from the theory. Thus, at the negative root
pole $\nab_-^2=-2\Lambda+ {\cal M}_1^2$, the residue is
\beq
-\frac{2}{3}-\frac{1}{3\left(1-4\Lambda\left(\tb-6\ta\right)\right)}\,,
\eeq
which is always $-1$ in the limit $\Lambda\to 0$, so that
\beq
A= G_N\,
\left[2{T^\prime}^{\mu\nu}\, {\cal P}^{-1} \,T_{\mu\nu}-
T'\,{\cal P}^{-1}\,T\right]\,,\label{massless}
\eeq
where ${\cal P}= \tb \left(\nab^2+1/\tb\right) \nab^2$.
While at the positive root pole $\nab_+^2=-\frac{4\Lambda}{3}-{\cal M}_2^2$,
the residue is
\beq
-\frac{2}{3}+\frac{2\Lambda}{9}\cdot \frac{\tb}{\left(1+8\Lambda\ta\right)}\,,
\eeq
which is always $-2/3$ in the limit $\Lambda\to 0$, so that
\beq
A= G_N\,
\left[2{T^\prime}^{\mu\nu}\, {\cal P}^{-1} \,T_{\mu\nu}-\frac{2}{3}\,
T'\,{\cal P}^{-1}\,T\right]\,.\label{massive}
\eeq
The results~(\ref{massless}) and~(\ref{massive}) are the well
known the vDVZ amplitudes for massless and massive spin-$2$ fields in the
flat space-time backgrounds. In particular, for $\tb \neq 0$, the
lowest mass of a localized
graviton in an $AdS_4$ space is quadratic one of $\Lambda$, i.e.,
$\delta M^2\propto\Lambda^2$. The amplitude~(\ref{massive}) remains the same
even when we take the limit $\tb=0$, except that ${\cal P}=\nab^2$. So a clear
message with the higher derivatives is that, even though the propagator is
found to have a single pole in the limit
$\Lambda\to 0$,~$\tb\to 0$, (i.e., ${\cal P}=1/\nab^2$), there are two
different propagators - one is massless (Eq.~(\ref{massless})) and another is
massive (Eq.~(\ref{massive})). This has been possible due to two different
root solutions for $\nab^2$. Obviously there is no spin-$2$ ghost
with $\tb=0$ and also the residue at $\nab^2=-4\Lambda/3$ becomes zero.

\section*{\it (C). Pole at  $(\tb+6\ta)\nab^2-1=0$}

In fact the amplitude~(\ref{amplitude}) is singular at
$\nab^2=(\tb+6\alpha)^{-1}$. The amplitude is singular at
$M_{PF}^2=2\Lambda/3$ in the massive gravity defined by the Pauli-Fierz
action~\cite{Porrati}.
This allows us to define $\M^2=-2\Lambda(\tb+6\alpha)\nab^2$, and the
above singularity is similar to the singularity at $M_{PF}^2=2\Lambda/3$.
However, the amplitude is smooth in the limit $0<\M^2<2\Lambda/3$, and a
smooth limit $\M^2/\Lambda\to 0$ is possible to be achieved since
$\M^2\sim \Lambda^2$. A more remarkable point is that when we set
$\ta=\tb=0$, so that the curvature squared terms are in Gauss-Bonnet
combination, neither there is any unphysical pole nor there is any ghost.
Nonetheless we recover correct amplitudes for both the massless and massive
gravitons by taking, respectively, $\tb=0,~\Lambda\to 0$, and
$\Lambda\to 0\,~\tb\to 0$ limits. This is a quite remarkable result.

\subsection{Flat space limit $\Lambda=0$}

We find useful to discuss some important features of the four-derivatives
gravitation by taking the flat space ($\Lambda=0$) limit. The
non-relativistic potential for arbitrary $\alpha, \beta,
\gamma$ takes the following form (taking $\kappa^2=16\pi G_4$)
~\cite{KStelle,Hans} (see references in~\cite{Hans} for earlier work)
\beq U(r)=G_4M_*
\left[-\frac{1}{r}-\frac{1}{3}\,\frac{e^{-m_0 r}}{r}+\frac{4}{3}\,
\frac{e^{-m_2 r}}{r}\right]\,, \label{potential} \eeq where \beq
m_0^2=-\frac{1}{2\kappa^2\left(\tb+3\ta\right)}\,, ~~~
m_2^2=\frac{1}{\kappa^2\tb}\,,\label{m0m2terms}
\eeq
and, $m_0,\,m_2$ are the mass terms for spin-$0$ and spin-$2$
particles. These massive modes would rise to give Yukawa-type
interactions. With the choice $\tb+3\ta=0\,,~|\tb\Lambda|<<1$, we find
\beq
\M_-^2=\frac{4\Lambda^2\tb}{3}+\frac{16\Lambda^3\tb^2}{3}\,, ~
\M_+^2=\frac{2\Lambda}{3}-\frac{16\Lambda^2\tb}{9}\,.
\eeq
In the limit $\tb\to 0$, $\M_+^2=2\Lambda/3$ and $\M_-^2=0$. Clearly,
when $\M_-^2$ represents mass of the spin-$2$ graviton of pure Einstein's
theory, $\M_+^2$ may characterize a mass term for the graviton,
but this also vanishes for $\Lambda=0$. In the
limit $\Lambda\to 0$, though both $\M_-^2$ and $\M_+^2$ approach zero,
$\M_-^2$ reaches to zero faster than $\M_+^2$. For $\tb\neq 0$,
the massive spin-$2$ field can behave as ghost-like. Another way
of seeing the presence of ghost-behavior is that one of the correction terms
to the Newtonian potential in~(\ref{potential}) has a coefficient $4/3$
with wrong (positive) sign, and an ultimate way to make the
theory free of ghost at the tree level is to set $\tb= 0$.
It is remarkable to note, which is also recently reviewed
in~\cite{Hans,Tekin}, that for finite $m_0,\, m_2$ and at large
distances, the potential~(\ref{potential}) reduces to Newtonian
limit, and it is finite for $r\to 0$. It was known that there is no
vDVZ discontinuity even with the Pauli-Fierz term when one sums
up all the tree-level graphs~\cite{Vainshtein}, i.e., in the full non-linear
analysis. Thus, with a non-zero cosmological constant, regardless one
introduces Pauli-Fierz term or higher derivatives, there is a
continuous limit for $M^2\to 0$ in the lowest order tree level
approximation, and we argue that this is a property of any space with
non-zero curvature.

From Eq.~(\ref{m0m2terms}) one finds
\beq
m_0^2+m_2^2=\frac{\tb+6\ta}{2\tb\left(\tb+3\ta\right)} \geq 0 \,.\nn
\eeq
When $\tb+6\ta=0$, one has $m_0^2=-m_2^2=1/\tb$. On the other hand,
in the same limit, one has $\M_-^2=\M_+^2=0=M^2$. Since $M^2=0$ limit
corresponds to the spin-$2$ massless graviton, we can therefore take the
limits $\ta\to 0\,, ~ \tb \to 0$ (i.e. $m_0\propto \infty,\,
m_2\propto \infty $) to arrive at the standard Newtonian limit for
non-relativistic potential.

A comment is in order. As stated above, for a non-zero $\tb$, there may
arise the problem of negative probabilities for processes
involving an odd number of massive spin-$2$ quanta and hence violation of
causality. However, since the parameters $\ta$ and $\tb$ are small enough to
make the massive fields, the breakdown of causality might occur only on a
microscopic scale or near a Planck scale. So a clear message is that if
the ghost are gentle and harmless creature, with whom we can live quite
comfortably, a choice $\tb\neq 0$, $\Lambda=0$ is possible, otherwise
we must have $\tb=0$. As we have seen above, ghost are gregarious which
demand the company of real particles, and they appear only at high energy,
so their effects will be insignificant in normal particle
scattering, which will appear to be unitary to a high degree of
accuracy~\cite{Hawking}.

\section{Gauge covariance and gauge fixing}

Any  viable gauge in a general four-derivative gravity theory
can involve third derivatives of the potentials (or metric fluctuations),
unlike the de Donder gauge in the Einstein-Hilbert theory, which involves only
first derivatives of the potential. In a flat space background $\Lambda=0$,
the linearized field equation~(\ref{neweqn}) is manifestly
invariant under the coordinate transformation
$x^\mu\longrightarrow {x^\p}^\mu= x^\mu + \xi^\mu$, where $\xi^\mu(x)$ is
infinitesimal vector field. For $\Lambda\neq 0$, however, the metric
fluctuations $h_{\mu\nu}$ are not, by themselves, invariant under
infinitesimal transformation. Here we shall make a cursory inspection
of gauge invariance for the $\Lambda\neq 0$ and $\Lambda=0$ cases.

\subsection{ A non-vanishing $\Lambda$}

To justify the above mentioned argument, let's write
the action in the following form (i.e., setting $\kappa^2=1,\,\beta=0,\,
\gamma=0$, so that $\tb=0$ and $\ta=\alpha$)
\beq
I_1 = \int d^4x \sqrt{- g} \left( R -2\Lambda + \alpha R^2 \right)
-2\int d^4x \sqrt{- g}\,h_{\mu\nu} T^{\mu\nu}\,.
\label{action1}
\eeq
For the background solutions~(\ref{Sol}), the equations linear in
$h_{\mu\nu}$ read
\beqa
&&{\cal L}_E(h_{\mu\nu})+2\Lambda h_{\mu\nu}-
4\Lambda^2\alpha\left(4 h_{\mu\nu}-
\bar{g}_{\mu\nu} h\right)+8\Lambda\alpha \Delta_L^{(2)}h_{\mu\nu}\nn \\
&&+4\Lambda\alpha \bar{g}_{\mu\nu}
\left(\nab^2h-\nab_\lambda\nab_\rho h^{\lambda\rho}\right)+
8\Lambda \alpha\left(2\nab_{(\mu}\nab^\lambda
h_{\nu)\lambda}-\nab_\mu\nab_\nu h\right)\nn \\
&&-4\alpha\left(\nab_\mu\nab_\nu-\bar{g}_{\mu\nu}\nab^2\right)
\left(-\nab^2h+\nab_\lambda\nab_\rho h^{\lambda\rho}-\Lambda h\right)
=4T_{\mu\nu}\,,\label{alphaeqn}
\eeqa
where ${\cal L}_E$ is the linearized Einstein term given by
\beq
\Delta^{(2)}_Lh_{\mu\nu} +
2\nab_{(\mu}\nab^\rho h_{\nu)\rho}
 -\nab_{(\mu}\nab_{\nu)}h -\Lambda
\big(4 h_{\mu\nu}-\g_{\mu\nu} h\big)
+\g_{\mu\nu}\left(\nab^2h -\nab_\lambda\nab_\rho
h^{\lambda\rho}\right)
\eeq
As is known, $\alpha=0$
explains only the spin-$2$ massless graviton, thus the
Eq.~(\ref{alphaeqn}) is clearly invariant under the gauge
transformation $h_{\mu\nu}\to h_{\mu\nu}+\nab_{(\mu} V_{\nu)}$. One can
use this invariance to set
\beq
\nab^\mu h_{\mu\nu}=\half\,\nab_\nu h\,,\label{gauge-choice}
\eeq
which removes four degrees of freedom. Set $T_{\mu\nu}=0$ and
$\alpha=0$ in Eq.~(\ref{alphaeqn}), find its trace, multiply it by
$g_{\mu\nu}/2$, and subtract the resultant expression from the
Eq.~(\ref{alphaeqn})
itself. Then one can use the gauge~(\ref{gauge-choice}) in order to arrive at
\beq
\Delta_L^{(2)} h_{\mu\nu}-2\Lambda h_{\mu\nu}=0\,.\label{aftergauge}
\eeq
The residual gauge invariance is then used to fix the above gauge,
which is solved by
\beq
\nab^\mu\nab_{(\mu} V_{\nu)}-\frac{1}{2}\,\nab_\nu \nab^\mu V_\mu
=\frac{1}{2}\left(\nab^2+\Lambda\right)V_\nu=0\,.\label{gauge-fixing}
\eeq
The Eq.~(\ref{aftergauge}) then implies
\beq
\nab_{(\mu}\left(\Delta_L^{(1)}-2\Lambda\right) V_{\nu)}
= -\nab_{(\mu}\left(\nab^2+\Lambda\right) V_{\nu)}=0\,.
\eeq
Since this equation is exactly solved by the gauge
fixing condition~(\ref{gauge-fixing}), there remains only $2\,(10-4-4)$
physical propagating degrees of freedom~\cite{Porrati}, which
correspond to a massless spin-$2$ graviton.

Now for arbitrary $\alpha$, using the same gauge transformations
as above, we find
\beqa
&&(1+8\Lambda\alpha)\nab_{(\mu}\left(\Delta_L^{(1)}-2\Lambda\right)V_{\nu)}
+2\alpha\nab_\mu\nab_\nu\left(\nab^2-\Lambda\right)\nab_\lambda
V^\lambda \nn\\&&~~~~~~~~~~~~~~ +\alpha
\bar{g}_{\mu\nu}\left(\nab^2+2\Lambda\right)
\left(\nab^2+2\Lambda\right)\nab_\lambda V^\lambda =0\,.
\eeqa
The gauge-fixing condition does not automatically solve this equation
unless $\alpha=0$ or $\nab_\lambda V^\lambda=0$. One finds then
three\footnote{Notice that there should be an extra constraint, so
that only $3$ out of the $4$ gauge variables are free.}
($=10-4-3\,,$ not just two!) physical degrees of freedom in the
theory defined by~(\ref{action1}) - two of them correspond to a
massless graviton and the rest one to a massive scalar excitation. B.~Whitt
in~\cite{Brian} showed that the fourth-order gravity
theory defined by~(\ref{action1}) is conformally equivalent to Einstein's
gravity with a massive scalar field, thus there is an extra scalar DOF in
the theory defined by~(\ref{action1}).

Further we consider the action by adding to Einstein-Hilbert action a
term quadratic in Ricci tensor
\beq
I_2 = \int d^4x \sqrt{- g} \left( R -2\Lambda + \beta R_{\mu\nu} R^{\mu\nu}
\right) -2\int d^4x \sqrt{- g}\,h_{\mu\nu} T^{\mu\nu}\,,
\label{action2}
\eeq
The linearized equations for this action take the following form, up to the
terms that vanish on Einstein backgrounds~(\ref{Sol}),
\beqa
&&\left(1-\beta\Delta_L^{(2)}+2\Lambda\beta\right)
\left(\nab_{(\mu}\left(\Delta_L^{(2)}-2\Lambda\right) V_{\nu)}\right)
-\beta\nab_\mu\nab_\nu\left(\Delta_L^{(0)}-2\Lambda\right)
\nab_\lambda V^\lambda\nn\\
&&-4\Lambda^2 \beta \nab_{(\mu} V_{\nu)}
+\frac{\beta}{2}\,\Delta_L^{(0)}
\left(\Delta_L^{(0)}-2\Lambda\right)\nab_\lambda V^\lambda
-\frac{2\Lambda^2\beta}{3}\, \bar{g}_{\mu\nu} \nab_\lambda
V^\lambda=0 \,.
\eeqa
Obviously, the gauge fixing condition~(\ref{gauge-fixing}) does not
solve this equation. What this implicitly implies is that the theory
defined by~(\ref{action}) has massive modes, other than the usual
massless spin-$2$ graviton, i.e., not all the components of $h_{\mu\nu}$ can
be fixed by a single gauge. It is known that the theory
defined by~(\ref{action2}) has a massive spin-$2$ mode other than the
usual massless graviton~\cite{KStelle,Odint}.

\subsection{The vanishing $\Lambda$}

In a flat space background, one may separate the fluctuations
$h_{\mu\nu}$ into massless spin-$2$ field\footnote{Note that
$h_{\mu\nu}^{(E)}$ represents the massless spin-$2$ field (graviton) of the
Einstein's theory when $\ta=\tb=0$.} $\phi_{\mu\nu}
(= h_{\mu\nu}^{(E)})$ and massive fields $\Sigma_{\mu\nu}$. The field
$\Sigma_{\mu\nu}$ can be broken up into a pure spin-$2$ field
$\sigma_{\mu\nu}$ and a massive scalar $\psi$. Being more precise
\beq
h_{\mu\nu}=\phi_{\mu\nu}+\Sigma_{\mu\nu}\,, ~~
\Sigma_{\mu\nu}=\sigma_{\mu\nu}+\eta_{\mu\nu} \psi
+2m^{-2}\partial_\mu\partial_\nu \psi\,.
\eeq
However, since we also have the massless field $\phi_{\mu\nu}$, the gradient
term above may be dropped, for it can be absorbed into a gauge transformation
of $\phi_{\mu\nu}$. Not surprisingly, the gauge invariance of the original
action is characterized by a gauge invariance for the $\phi_{\mu\nu}$ field
alone.

With $\Lambda=0$, the linear equations derived in ~(\ref{neweqn}) take the
following form
\beq
\left(1+\tb \Box \right) \left(\delta R_{\mu\nu}- \frac{1}{2}\,\eta_{\mu\nu}
\delta R\right)- (\tb+2\ta)
\left(\partial_\mu\partial_\nu-\eta_{\mu\nu}\Box\right)\delta R=
2 T_{\mu\nu}\,,\label{flateqn}
\eeq
where
\beqa
\delta R_{\mu\nu}&=&-\frac{1}{2}\,\Box h_{\mu\nu}+\partial_{(\mu}\partial^\rho
\widetilde{h}_{\nu)\rho}\,,~~
\delta R=-\frac{1}{2}\,\Box h +\partial^\mu\partial^\nu \wt{h}_{\mu\nu}\,,\nn\\
\Box&=&\eta_{\mu\nu} \partial^\mu\partial^\nu\,,~~~
\widetilde{h}_{\mu\nu}=h_{\mu\nu}-\frac{1}{2}\,\eta_{\mu\nu}
h\,.\label{formula}
\eeqa
The trace of Eq.~(\ref{flateqn}) gives
\beq
\frac{1}{3}\,\left(1+\tb \Box\right)\delta R -
\left(\tb+2\ta\right) \Box\,\delta R= -\frac{2}{3}\,
T\,.\label{newtrace}
\eeq
Then from the Eqs.~(\ref{flateqn})
and~(\ref{newtrace}), by using~(\ref{formula}), we arrive to
\beq
\left(1+\tb
\Box\right)\left[-\frac{1}{2}\,\Box h_{\mu\nu}-\frac{1}{6}\,
\eta_{\mu\nu} \delta R\right] + \partial_{(\mu}\Gamma_{\nu)}=
2\left(T_{\mu\nu}-\frac{1}{3}\,\eta_{\mu\nu}T\right)\,,\label{linear4}
\eeq
where
\beq
\Gamma_\mu\equiv \left( 1+\tb \Box\right)
\partial^\rho \wt{h}_{\mu\rho} -\left( \tb + 2\ta\right)
\partial_\mu\, \delta R\,.\label{Teyssendier}
\eeq
As expected, for $\ta=\tb=0$, the field equations~(\ref{linear4})
and~(\ref{newtrace}) in the gauge $\partial^\rho
\wt{h}^{(E)}_{\mu\rho}=0$ yield
\beq
\Box\,\phi_{\mu\nu}=-4\left(T_{\mu\nu}-\frac{1}{2}\,\eta_{\mu\nu}
T\right)\,, \label{Einsteinpart}
\eeq
where $h_{\mu\nu}^{(E)}$ is
represented by $\phi_{\mu\nu}$. Notice that under the coordinate
transformation $x^\mu\longrightarrow {x^\p}^\mu= x^\mu + \xi^\mu$,
the variables $\Gamma_\mu (x)$ transform as
\beq
\Gamma_\mu(x)\longrightarrow \Gamma^\p_\mu(x)=\Gamma_\mu (x) -
\left(1+\tb \Box\right)\Box\, \xi_\mu(x)\,.
\eeq
Now the gauge must be fixed by choosing some arbitrary relations which
determine the $\xi_\mu$ in terms of arbitrary variables and their
derivatives. Practically, one can also choose a gauge that
determines $\xi_\mu$ in terms of the given Cauchy data on the
initial hypersurface. A simple viable gauge is $ \Gamma_\mu(x)=0$
, which is known as ``Teyssandier'' gauge~\cite{Teyssandier}, and
this gauge choice was justified in Ref.~\cite{Accioly1} by choosing
$\gamma=0$. Obviously, for $\ta=\tb=0$, Teyssandier gauge reduces to de Donder
gauge $\partial^\rho \tilde{h}_{\mu\rho}=0$. With $\Gamma_\mu=0$,
the Eq.~(\ref{linear4}) can be written as
\beq
\left(\Box
+m_2^2\right)\sigma_{\mu\nu}= 4\left(T_{\mu\nu}-\frac{1}{3}\,
\eta_{\mu\nu} T\right)\,,\label{sigmaeqn}
\eeq
where $m_2^2=1/\tb$ has been used, and the massive spin-$2$ field
$\sigma_{\mu\nu}$ is defined by
\beq
\sigma_{\mu\nu}\equiv \frac{-\Box
h_{\mu\nu}-(1/3)\, \eta_{\mu\nu}\delta
R}{m_2^2}\,.\label{sigmavalue}
\eeq
Indeed, consistency of
Eqs.~(\ref{sigmaeqn}) and~(\ref{sigmavalue}) with $\Gamma_\mu=0$,
latter as a gauge condition, also requires that
$\partial^\mu\partial^\nu \sigma_{\mu\nu}-\Box \sigma=0$, which is
the reminiscent of the constraint variables $\nab^\mu\nab^\nu
h_{\mu\nu}-\nab^2 h=0$ in the full theory defined
by~(\ref{neweqn}). Eq.~(\ref{sigmaeqn}) may be written as, using
Eq.~(\ref{sigmavalue}),
\beq
\Box \sigma_{\mu\nu}-\Box h_{\mu\nu}-\frac{1}{3}\,\eta_{\mu\nu}
\delta R= 4\left(T_{\mu\nu}-\frac{1}{3}\, \eta_{\mu\nu}
T\right)\,.\label{sigmaeqn2}
\eeq
However, in the gauge
$\Gamma_\mu=0$, the trace part of~(\ref{linear4}) yields
\beq
-\frac{1}{2}\,\left(1+\tb \Box\right)\Box h-\frac{2}{3}\left(1+\tb
\Box\right) \delta R=-\frac{2}{3}\, T\,,\label{flattrace1}
\eeq
while the gauge $\Gamma_\mu=0$ itself implies
\beq
\frac{1}{2}\,\left(1+\tb \Box\right)\Box h +\left(1+\tb
\Box\right)\delta R -\left(\big(\tb+2\ta\big)\Box\right)\delta
R=0\,.\label{gaugetrace1}
\eeq
Hence the Eqs.(\ref{flattrace1})
and~(\ref{gaugetrace1}) would rise to give
\beq
\left[1-2\left(\tb+3\ta\right)\Box\right] \delta R =-2
T\label{m0eqn}\,.
\eeq
Substitute this value of $\delta R$ into the
Eq.~(\ref{sigmaeqn2}) and arrive to
\beq
\Box\left(h_{\mu\nu}-\sigma_{\mu\nu}-\frac{1}{3
m_0^2}\,\eta_{\mu\nu}\, \delta
R\right)=-4\left(T_{\mu\nu}-\frac{1}{2}\,\eta_{\mu\nu} T\right)\,.
\label{massiveparts}
\eeq
where as defined previously
$m_0^2=-1/\big(2(\tb+3\ta)\big)$. Evidently,
Eqs.~(\ref{Einsteinpart}) and~(\ref{massiveparts}) reveal that
\beq
h_{\mu\nu}=\phi_{\mu\nu}+\sigma_{\mu\nu}+\eta_{\mu\nu} \psi\,,
\eeq
where the massive scalar field $\psi=\delta R/(3m_0^2)$
satisfies the following equation (from Eq.~(\ref{m0eqn}))
\beq
\left(\Box+m_0^2\right) \psi = -\frac{2}{3}\, T\,.
\eeq
Thus, in the momentum space, the interaction amplitude between two
conserved sources ${T^\p}^{\mu\nu}$ and $T_{\mu\nu}$ is given by
\beq
\frac{1}{2}\,h_{\mu\nu} {T^\p}^{\mu\nu}= - 8\pi G_4
\left[\frac{\left(T_{\mu\nu}{T^\p}^{\mu\nu}-\frac{1}{2}\,
T_\lambda^\lambda {T^\p}_\rho^\rho\right)}{p^2}
- \frac{\left(T_{\mu\nu}{T^\p}^{\mu\nu}-\frac{1}{3}\,T_\lambda^\lambda
{T^\p}_\rho^\rho\right)}{p^2+m_2^2}
+\frac{T_\lambda^\lambda {T^\p}_\rho^\rho}{6\left(p^2+m_0^2\right)}\right]\,.
\eeq
Thus to make the theory free from a massive spin 2 ghost state, we set
$\tb=0$, so that the second term above is formally absent.

\subsection{Gauge-fixing term}

The gauge invariance of~(\ref{flateqn}) is made manifest by
writing the action in the following bilinear form
\beq S=\int d^4x
\left[\frac{1}{4}\,h^{\mu\nu}\,{\cal K}_{\mu\nu\rho\sigma}^{inv}\,
h^{\rho\sigma}+2h^{\mu\nu}\, T_{\mu\nu}\right]\,,\label{actiong}
\eeq
where the differential operator kernel ${\cal K}^{inv}$ for
the diff-invariant part is
\beq {\cal K}^{inv}= \Box (1+\tb
\Box)\, P^{(2)}_{\mu\nu\rho\sigma}
-2\Box\left(1-2(\tb+3\ta)\Box\right)\,P^{(S)}_{\mu\nu\rho\sigma}\,,
\eeq
where the transverse spin-$2$ and spin-$0$ (scalar) projectors are
defined by
\beq
P^{(2)}_{\mu\nu\rho\sigma}=\frac{1}{2}\,\theta_{\mu\rho}\theta_{\nu\sigma}
+\frac{1}{2}\,\theta_{\mu\sigma}\theta_{\nu\rho}- \frac{1}{3}\,
\theta_{\mu\nu}\theta_{\rho\sigma}\,, ~~  P^{(S)}=\frac{1}{3}\,
\theta_{\mu\nu}\theta_{\rho\sigma}\,; ~~
\eta_{\mu\nu}-\theta_{\mu\nu}=
\frac{\partial_\mu\partial_\nu}{\Box}\equiv \omega_{\mu\nu}\,.
\eeq
Here $P^{(2)}$ and $P^{(S)}$ are symmetric under
$\mu\leftrightarrow\nu$, $\rho\leftrightarrow\sigma$,
$\mu\nu\leftrightarrow\rho\sigma$. As we illustrated in the last
subsection, in the absence of gauge fixing term, $K^{inv}$ explains
eight physical propagating degrees of freedom in the spin subspace
$2\oplus S$, which correspond to the massless graviton
$(\phi_{\mu\nu})$, the massive spin-$2$ field $(\sigma_{\mu\nu})$
and a massive scalar ($\psi$). Indeed, we can make the
four-derivative gravitational action complete by including
to~(\ref{actiong}) the following gauge fixing
term~\cite{Bartoli}~\footnote{As several authors have studied the
fourth-order gravity action with $\gamma=0$ supplemented by gauge fixing term
(see Ref.~\cite{Odint} for review), we will be brief here. A theory
defined with $\gamma\neq 0$ is more physical
in a sense that one obtains ghost free action with the choice
$\beta+4\gamma=0$, but in the formal case one must set $\beta=0$.}
\beq
S_{gf}=\int d^4x\, \chi^\mu(h)\, \G_{\mu\nu}\, \chi^\nu(h)\,,
\eeq where \beq \chi^\mu(h)\equiv \partial_\nu
h^{\mu\nu}-\lambda_1 \nab^\mu h\,, ~~ \G_{\mu\nu}\equiv \lambda_2
\nab^\rho\nab_\rho g_{\mu\nu}
+\frac{\lambda_3}{2}\,\nab_{(\mu}\nab_{\nu)}+\lambda_4\,
g_{\mu\nu}\,,
\eeq
with $\lambda_1,\cdots, \lambda_4$ being the
gauge fixing parameters. The gauge fixing contribution in terms of
the differential operator kernel is
\beq {\cal K}^{gf}=
-6\lambda_1^2\,\Box\lambda^*\,P^{(S)}-\Box \left(\lambda_4+\Box
\lambda_2\right)\,P^{(1)}-2(1-\lambda_1)
\left((1-\lambda_1)\Box \lambda^*\,P^{(W)} -\lambda_1\Box
\lambda^*\,P^{(SW)}\right)
\eeq
where $\lambda^*=\lambda_4+(\lambda_2+\lambda_3)\Box$, and the spin-$1$
operator $P^{(1)}$ and the transfer operators $P^{(W)}$,
$P^{(SW)}$ are \beq
P^{(1)}=\frac{1}{2}\,\theta_{\mu\rho}\omega_{\nu\sigma}+\frac{1}{2}\,
\theta_{\mu\sigma}\omega_{\nu\rho}\,, ~~
P^{(W)}=\omega_{\mu\nu}\omega_{\rho\sigma}\,,  ~~ P^{(SW)}
=\theta_{\mu\nu}\omega_{\rho\sigma}+\theta_{\rho\sigma}\omega_{\mu\nu}\,.
\eeq
The total contribution to the four-derivative differential
operator is given by the sum ${\cal K}={\cal K}^{inv}+{\cal K}^{gf}$.
One can redefine the field $h_{\mu\nu}$ by the following
transformation~\cite{Bartoli}
\beq
{\cal K}\longrightarrow {\widehat{\cal K}}= A\,{\cal K}\,A\,,
\eeq
so that the operator $A(\lambda_1)$ defined by
\beq
A(\lambda_1)=P^{(2)}+P^{(1)}+\frac{2}{3}\,P^{(W)}+\frac{2}{9}\,
\frac{(1-\lambda_1)}{\lambda_1}\,P^{(SW)}
\eeq
is invertible, and becomes a numerical matrix for $\lambda_1=-2$. While, the
new field $\hat{h}_{\mu\nu}$ is transformed as
\beq \hat{h}_{\mu\nu}=
(A^{-1})_{\mu\nu}^{\rho\sigma}\,h_{\rho\sigma}
\eeq
Finally, the
quartic propagator obtained by inverting the projectors take the
following form~\cite{Bartoli} \beqa {\cal P}(\hat{h}) &=&
\frac{1}{\big(1+\tb \Box\big)\Box}\,P^{(2)}
-\frac{27}{8}\,\frac{\lambda_1^2}{(1-\lambda_1)^2\Big(1-2\big(\tb+3\ta\big)
\Box\Big)\Box}\,P^{(W)}\nn\\
&{}& -\frac{1}{\left(\lambda_4+\lambda_2\Box\right)\Box}\,P^{(1)}
-\frac{27}{8}\,\frac{\lambda_1^2}{(1-\lambda_1)^4\Big(\lambda_4
+\big(\lambda_2+\lambda_3\big)\Box\Big)}\,P^{(S)}\,.
\eeqa
This can be split into two new fields $\wwh$ and $\wwp$ as
\beqa
{\cal P}_1(\wwh)& =& \frac{1}{\Box}\,P^{(2)}
-\frac{27}{8}\,\frac{\lambda_1^2}{(1-\lambda_1)^2\,\Box}\,P^{(W)}
-\frac{1}{\lambda_4\,\Box}\,P^{(1)}
-\frac{27}{8}\,\frac{\lambda_1^2}{(1-\lambda_1)^4\,\lambda_4\,\Box}\,
P^{(S)}\,,\label{tildeh}\\
{\cal P}_2(\wwp) &=& - \frac{1}{\big(\Box+m_2^2\big)}\,P^{(2)}
+\frac{27}{8}\,\frac{\lambda_1^2}{(1-\lambda_1)^2}\,
\frac{1}{\Big(\Box+m_w^2\Big)}\,P^{(W)}\nn\\
&{}& +\frac{1}{\lambda_4\left(\Box+m_1^2\right)}
\,P^{(1)}+\frac{27}{8}\,\frac{\lambda_1^2}{(1-\lambda_1)^4}\,
\frac{1}{\lambda_4\Big(\Box+m_s^2\Big)}\, P^{(S)}\,.\label{tildep}
\eeqa where \beq m_2^2=\frac{1}{\tb}\,, ~~
m_w^2=-\frac{1}{2\big(\tb+3\ta\big)}\,, ~~
m_1^2=\frac{\lambda_4}{\lambda_2}\,, ~~ m_s^2=\frac{\lambda_4}
{\big(\lambda_2+\lambda_3\big)}\,.
\eeq
The sign in front of each spin operator in
Eqs.~(\ref{tildeh}) and~(\ref{tildep}) is important. Obviously, in
the massless $\wwh$ sector Eq.~(\ref{tildeh}), there are just two
physical propagating degrees of freedom in $P^{(2)}$, which
correspond to the massless graviton, while other three ghost modes
(if $\lambda_4>0$) are just gauge degrees. On the other hand,
the massive $\wwp$ sector (Eq.~(\ref{tildep})) contains a massive
spin-$2$ field of mass $m_2^2$ with five degrees of freedom and
one physical scalar with mass equal to $m_0^2$ defined previously,
and rest are just the gauge degrees of freedom, but they are not
necessarily ghost excitations, it depends upon the sign of gauge
parameter $\lambda_4$. If $\lambda_4<0$ (provided also that
$\lambda_2<0$ and $|\lambda_2|>|\lambda_3|$), all gauge dependent
massive fields in $\wwp$ sector become real gauge dependent fields,
otherwise they show ghost behavior. When we set $\tb=0$ and choose a
gauge such that $\lambda_1=0$, we will be left only with a massless
graviton as a physical spin 2 excitation of the Einstein's gravity.

It is essential to quantize the theory in order to extract any concrete
ideas about the vDVZ discontinuity at the quantum level. Obviously, one must
then add the compensating higher derivative Faddeev-Popov Lagrangian
including fermion ghosts and auxiliary commuting fields, this could be
an interesting topic for a separate publication.

\section{Conclusions}

In conclusion, we have seen that the leading order curvature terms as
quadratic corrections to Einstein's gravity define a theory with a
number of striking properties. The absence of van Dam-Veltman-Zakharov
discontinuity in $AdS$ space in the $M^2 \to 0$ limit of the
massive graviton is just a simple
example shown above. Though one may have to check whether these results
persist up to the loop level, such effects are much
suppressed and would be insensitive to the present experimental
observations, like gravitational bending. Nonetheless, one can expect
vDVZ discontinuity in the
graviton propagator to be absent even at loop level since the mass
of the graviton is generated dynamically from the higher curvature
terms. There is a clear message that a quadratic gravity itself
deserves as a consistent theory of massive spin-$2$ field with
sensible massless limit. This is the main thrust of this paper. We
also obtained the general solutions to the linearized higher
derivative field equations by specializing to the $\Lambda=0$ case
in the Teyssandier gauge and explicitly identified the physical
eight degrees of freedom of the theory. Finally we completed our
discussion with a cursory inspection of the general gauge
covariance by including the gauge fixing term.

We argue that in any effective $4D$ theory
obtained from the dimensional reduction of higher dimensional
theory both $M^2$ and $H_0^2$ ($H_0$ is the Hubble parameter) could
depend on effective $4D$ cosmological constant. Since $\Lambda\sim
H_0^2$, the ratio $M^2/\Lambda\sim M^2/H_0^2$ when remains finite, a
massive graviton may contribute to gravity with a meaningful
phenomenology. It is probable that a small fraction of the gravitational
interactions is associated with a massive graviton component, which could
be even ghost-like, while still dominant is the component of massless
gravitons, and this is indeed a general feature of four-derivatives
gravitation. In brane world models, on the other hand, one may realize
similar effects by having an ultra-light spin-$2$ particle with a very small
coupling compared to massless graviton as pointed out
in~\cite{KR,Dvali,IIK}.

\section*{Acknowledgments}

It is a pleasure to thank Gregory Gabadadze, Andreas Karch, Ian Kogan and
Antonios Papazoglou for fruitful correspondences and important comments in
the earlier version of this manuscript, and Hyun Min Lee and Y. S. Myung
for helpful discussions. The author
acknowledges a Seoam Fellowship, and this work was partly supported by the
Brain Korea 21 Program, Ministry of Education.

\section*{Appendix A: Linear Field Equations for $R^2$-Gravity}
\renewcommand{\theequation}{A.\arabic{equation}}
\setcounter{equation}{0}


For the mathematical clarity, we give the following variational formulae for
different curvature and derivative terms.

(i) Variation of $G_{\mu\nu}$
\beqa
\delta G_{\mu\nu}&=&\delta R_{\mu\nu}-\frac{1}{2}\, \bar{g}_{\mu\nu}\delta R
-\half h_{\mu\nu}
\R \Longrightarrow \delta R_{\mu\nu}= \delta G_{\mu\nu}+\frac{1}{2}\,
\bar{g}_{\mu\nu}
\delta R+\half\, h_{\mu\nu}\R\nn \\
\delta G_{\mu\nu}&=& \half\,\left(\Delta^{(2)}_Lh_{\mu\nu}
+2 \nab_{(\mu}\nab^\rho h_{\nu)\rho}
 -\nab_{(\mu}\nab_{\nu)}h -\Lambda\,
\big(4 h_{\mu\nu}-\g_{\mu\nu} h\big)
+\g_{\mu\nu}\left(\nab^2h -\nab_\lambda\nab_\rho
h^{\lambda\rho}\right)\right)\nn \\
&& ~~
\eeqa

(ii) Variation of $I_{\mu\nu}\left(\equiv R\left(R_{\mu\nu}-\frac{1}{4}\,
g_{\mu\nu}R\right)\right)$
\beqa
\delta I_{\mu\nu}&=& \left(\R_{\mu\nu}-\half\,\g_{\mu\nu}\R\right)
\delta R +\R \delta R_{\mu\nu}-\frac{1}{4}\, h_{\mu\nu} \R^2 \nn\\
&=& \Lambda \g_{\mu\nu}\left(\nab^2h-\nab^\lambda\nab^\rho h_{\lambda\rho}
\right) +2\Lambda\left(\Delta_L^{(2)}+2\nab_{(\mu}\nab^\lambda
h_{\nu)\lambda}-\nab_\mu\nab_\nu h\right)
-4\Lambda^2\left(h_{\mu\nu}-\frac{1}{4}\,\g_{\mu\nu} h\right)\nn \\
&& ~~
\eeqa

(iii) Variation of $J_{\mu\nu}\left(\equiv R_{\mu\rho\nu\sigma} R^{\rho\sigma}
-\frac{1}{4}\,g_{\mu\nu} R_{\rho\sigma} R^{\rho\sigma}\right)$

\beqa
\delta J_{\mu\nu}&=& \delta R_{\mu\rho\nu}\,^\sigma \R^\rho\,_\sigma
+\R_{\mu\rho\nu\sigma}\delta R^{\rho\sigma}
-\half\, \bar{g}_{\mu\nu}\R^{\rho\sigma}\delta R_{\rho\sigma}\nn \\
&{}&
-\left(\R_{\mu\rho\nu}\,^\lambda \R_{\sigma\lambda}-\half\,\bar{g}_{\mu\nu}
\R_{\rho\lambda}\R_\sigma\,^\lambda\right) h^{\sigma\rho}
-\frac{1}{4}\, h_{\mu\nu}\R_{\rho\sigma}\R^{\rho\sigma}\nn \\
&=&\frac{\Lambda}{2}\left(\Delta_L^{(2)} h_{\mu\nu}+2\nab_{(\mu}\nab^\lambda
h_{\nu)\lambda}-\nab_\mu\nab_\nu h\right)-2\Lambda^2 \left(h_{\mu\nu}
-\frac{1}{4}\,\g_{\mu\nu} h\right)\nn \\
&{}& -\frac{\Lambda}{3}\,\delta R_{\mu\nu}
+\frac{\Lambda}{6}\,\g_{\mu\nu}\left(\nab^2h-\nab^\lambda\nab^\rho
h_{\lambda\rho}\right)
\eeqa

(iv) Variation of $K_{\mu\nu} \left(\equiv R_{\mu\rho\sigma\delta}
R_\nu\,^{\rho\sigma\delta}
-\frac{1}{4}\,g_{\mu\nu} R_{\rho\sigma\delta\lambda}
R^{\rho\sigma\delta\lambda} -2 R_\mu\,^\lambda R_{\nu\lambda}
+2R_{\mu\lambda\nu\rho} R^{\lambda\rho}\right)$

\beqa
\delta K_{\mu\nu}&=& 2\R_{(\mu}\,^{\rho\sigma}{}_{|\lambda|}\,\delta R_{\nu)
\rho\sigma}\,^\lambda-\half\, \bar{g}_{\mu\nu}
\R^{\rho\sigma\lambda}\,_\tau \delta R_{\rho\sigma\lambda}\,^\tau
+2\R^\rho\,_\sigma\delta R_{\mu\rho\nu}\,^\sigma+2\R_{\mu\rho\nu\sigma}\delta
R^{\rho\sigma}-4\R_{(\mu}\,^\rho \delta R_{\nu)\rho}\nn \\
&{}&-\Big(\R_{\mu\rho\alpha\beta} \R_{\nu\sigma}\,^{\alpha\beta}
-\half\,\bar{g}_{\mu\nu}\R_{\rho\alpha\beta\lambda}\R_\sigma\,^{\alpha\beta\lambda}+2\R_{\mu\rho\nu}\,^\lambda \R_{\sigma\lambda}
-2\R_{\mu\rho}\R_{\nu\sigma}\Big)h^{\rho\sigma}-\frac{1}{4}\,
h_{\mu\nu}\R_{\alpha\beta\lambda\tau}\R^{\alpha\beta\lambda\tau}\nn \\
&& ~~
\eeqa

(v) Variation of $\nabla_\mu\nabla_\nu R$
\beq
\delta \left(\nabla_\mu\nabla_\nu R\right)
=\nab_\mu\nab_\nu \delta R-\nab_\lambda R\,
\delta\Gamma_{\mu\nu}\,^{\lambda}
\eeq
where
\beq
\delta\Gamma_{\mu\nu}\,^{\lambda}=\frac{1}{2}\,\g^{\lambda\sigma}
\left(\nab_\mu h_{\nu\sigma}+\nab_\nu h_{\mu\sigma}-\nab_\sigma h_{\mu\nu}
\right)
\eeq

(vi) Variation of $g_{\mu\nu}\nabla^2 R$
\beq
\delta\left(g_{\mu\nu}\nabla^2 R\right)
=\nab^2 R h_{\mu\nu}-\g_{\mu\nu} h^{\lambda\rho}
\nab_\lambda\nab_\rho \R
+\g_{\mu\nu}\left(\nab^2\delta R-\g^{\rho\sigma}
\delta\Gamma_{\rho\sigma}\,^\lambda\,\nab_\lambda\R\right)
\eeq

(vii) Variation of $\nabla^2 R_{\mu\nu}$

\beq
\delta\left(\nabla^2 R_{\mu\nu}\right)
=\nab^2\delta R_{\mu\nu}-h^{\rho\sigma}\nab_\rho\nab_\sigma
\R_{\mu\nu}-g^{\rho\sigma}
\delta\Gamma_{\rho\sigma}\,^\lambda\,\nab_\lambda\R_{\mu\nu}
-\left(2\nab^\rho\R_{\mu\sigma}\delta\Gamma_{\nu\rho}\,^\sigma
+\R_{\mu\sigma}\nab^\rho\delta\Gamma_{\nu\rho}\,^\sigma
+\mu\leftrightarrow\nu\right)
\eeq

Eq.~(\ref{eq.of.mo}) can also be written in the following form
($\kappa^2=1$)
\beqa
&&\left(R_{\mn}-\fr{1}{2}g_{\mn}R\right)+
\Lambda g_{\mu\nu}+2\ta R\left(R_{\mu\nu}-\frac{1}{4}
g_{\mu\nu} R\right)+2\tb\left(R_{\mu\rho\nu\sigma} R^{\rho\sigma}
-\frac{1}{4}g_{\mu\nu} R_{\rho\sigma}R^{\rho\sigma}\right)\nn \\
&&~~~~~~~+2\gamma \left(R R_{\mu\nu}-2R_{\mu\rho\nu\sigma}R^{\rho\sigma}+
R_{\mu\rho\sigma\lambda}R_\nu\,^{\rho\sigma\lambda}-2R_\mu\,^\rho
R_{\nu\rho}-\frac{1}{4}\,g_{\mu\nu} {\cal R}_{GB}^2\right)\nn\\
&& ~~~~~~~~~~~~-\left(\tb+2\ta\right)\left(\nabla_\mu
\nabla_\nu-g_{\mu\nu} \nabla^2\right)R +\tb\,
\nabla^2\left(R_{\mu\nu}-\frac{1}{2}\, g_{\mu\nu} R\right)=0\,,
\label{eq.of.mo2}
\eeqa
where, as defined in the text, $\ta=\alpha-\gamma$,\, $\tb=\beta+4\gamma$,
and the Gauss-Bonnet curvature squared ${\cal R}_{GB}^2=
R^2-4R_{\rho\sigma}^2+R_{\rho\sigma\lambda\tau}^2$. For the
background solutions~(\ref{Sol}), terms in the bracket multiplying with
$2\gamma$ have a trivial contribution for the linear perturbations. While,
on the Einstein backgrounds~(\ref{Sol}), the non-trivial contribution from
the last two terms that involve covariant derivatives is
\beq
-\left(\tb+2\ta\right)\left(\nab_\mu\nab_\nu-\g_{\mu\nu}\nab^2\right)
\delta R+\tb\, \nab^2\delta G_{\mu\nu}+2\Lambda\tb\, \nab^2 h_{\mu\nu}\,.
\label{fourth-order}
\eeq
Finally, using the results~(i)-(iii), and~(\ref{fourth-order}), we arrive to
Eq.~(\ref{neweqn}) given in the text.

\end{document}